\documentclass[preprint]{revtex4}
\usepackage{hyperref}
\usepackage{graphicx}
\begin{document}
\title{ Mirage Cosmology of $U(1)$ Gauge Field on Unstable D3 Brane Universe}
\author{Anastasios Psinas}
\affiliation{Department of Physics,\\ Northeastern University \\
Boston, MA 02115-5000, USA} \email{psinas.a@neu.edu}
\begin{abstract}
An unstable $D3$-brane universe governed by the DBI action of the
tachyon field minimally coupled to a $U(1)$ gauge boson is
examined. The cosmological evolution of this coupled system, is
further analyzed, in terms of the expansion rate of the inflating
brane, which is highly affected by the presence of the tachyonic
and gauge field charges. We show, that the minimal coupling makes
the effective brane density less divergent. However, for some
sectors of the theory the tachyon is not able to regulate it in an
efficient fashion. Also, a detailed analysis of the dependance of
the effective brane density on the scale factor of the universe is
performed, which leads to various cosmological models.
\end{abstract}
%\keywords{D-Branes, $U(1)$, Gauge fields, Cosmology, String
%theory} \pacs{71837189 } \maketitle
%\pacs{04.50.+h, 98.80.Cq, 11.10Kk}
\maketitle

\section{Introduction}
  There has been remarkable progress since the seminal work of
Guth \cite{Guth:1980zm} which was one of the first works to
incorporate particle physics in cosmological analyses. After the
emergence of string theory, the missing link was to find a
consistent way to accommodate stringy physics in cosmology. Among
the series of models that attempted to fill this gap, brane gases
emerged as a promising approach in terms of T-duality implemented
in cosmology. Thus, in these models, one can study the case where
our universe, is filled with a thermal bath of strings or even
M-theory membranes such that the compact dimensions of the
universe can be wrapped around branes. So in this fashion the
winding numbers affect the cosmological evolution
\cite{Alexander:2000xv,Brandenberger:2001kj,Brandenberger:1988aj,Tseytlin:1991xk}.

 Quite recently
 \cite{Sen:2002nu,Garousi:2000tr}, it was shown that
string theory can accommodate unstable Dp-brane configurations
apart from the already known stable ones. Following along the same
lines an effective action for non-BPS D-branes in Type II string
theories was proposed, where the manifested instability is caused
by the presence of the tachyon field. Roughly speaking, the action
governing this system is still the Dirac-Born-Infeld action
modified in a way, that after tachyon condensetation takes place
we are left with stable D-branes. To this end, it is necessary,
that one reverts to a decaying like tachyonic potential
\cite{Kutasov:2000qp,Kutasov:2004dj}. Moreover, tachyon physics
provided us with a deeper understanding regarding cosmological
inflation
\cite{Frolov:2002rr,Kofman:2002rh,Felder:2002sv,Shiu:2002qe,Gibbons:2002md,Feinstein:2002aj,Padmanabhan:2002sh,Kim:2002zr,Shiu:2002xp,Gorini:2004by,Piao:2003sc,Panigrahi:2004qr}.
In the very elucidating review \cite{Gibbons:2003gb} tachyon
cosmology was also examined in terms of a "No-Go theorem" of
warped compactifications \cite{Gibbons:1984kp}. For a more
extensive analysis on this issue see \cite{Maldacena:2000yy} and
references therein.

However, several years ago, a different pathway was followed
\cite{Randall:1999ee,Randall:1999vf}. Yet again, the inspiration
came form $11$ -dim supergravity \cite{Horava:1996ma} where
instead of dealing with M-theory on $R^{10} \times{ S^{1}/Z_{2}}$
the authors compactified a five dimensional AdS spacetime on
$S^{1}/Z_{2} $. The salient trait of this model, was that our $4$
dimensional observed universe is fixed in one of the two
hyperplanes of the orbifolded spacetime discribed above and that
gravity is free to propagate in the bulk but the graviton is
actually confined in our universe since its wave function decays
exponentially as it tries to expand away from the visible brane.
Of course, one can consider apart from warped compactification
schemes as in RS model, the case of extra open dimensions (
addressing the hierarchy problem as well) as it is done in
\cite{Arkani-Hamed:1998rs,Antoniadis:1998ig}. Also, a relevant
extension (non-supersymmetric) based on  the Intersecting Brane
World scenarios has been found \cite{Cremades:2002dh}.

If one seeks  more concrete cosmological models in terms of
D-brane physics, then it is useful to resort to brane world
scenarios that are inspired from the well celebrated AdS/CFT
correspondence \cite{Maldacena:1997re}. An elegant cosmological
construction based on this correspondence, was first introduced by
Kehagias-Kiritsis \cite{Kehagias:1999vr} and became known as
mirage cosmology. In this context, we have the notion of a
D$3$-brane representing our $4$-dim universe that is embedded in a
higher dimensional ambient space filled with a stack of Dp-branes
with p$>3$. The core of this approach, is based upon the fact,
that the D$3$-brane probe follows a geodesic motion under the
influence of the gravitational field generated by the stack of
Dp-branes. It is also considered that the stack is much heavier
than the probe so one can safely ignore the back-reaction. What is
really interesting about this realization, is the fact that the
motion of the D$3$-brane induces a cosmological evolution, even in
the case where our universe is empty of matter. Therefore, any
cosmological expansion or contraction of our visible universe can
be attributed to energy density that can originate from anywhere
else other than the probe brane such us the bulk. Due to the
intrinsic richness of this model, a thorough analysis of its
implications in various aspects of string and D-brane inspired
cosmologies followed
\cite{Papantonopoulos:2000xs,Youm:2000mi,Youm:2000ke,Brax:2002qw,Bozhilov:2002fa,Kim:2002hk,Piao:2002vf,Boehm:2002kf,Papantonopoulos:2004au,Iwashita:2005wi}.

One in principle, can also study mirage cosmology when considering
tachyonic degrees of freedom living in the bulk
\cite{Kim:2000ax,Papantonopoulos:2000wv,Diamandis:2002zn}. While
the basic aspects of these models will be discussed in the later
part of this paper, it is worthwhile to mention, that other
extensions can also be considered \cite{Jeong:2005py}. In fact,
our whole analysis is strongly influenced by this paper in which
the tachyon field is constrained on the D$3$-brane probe. What we
propose, is to examine the case where one also turns a gauge field
on the brane which is minimally coupled to the tachyon. For
completeness, we mention, that the study of winding tachyon
coupled to a gauge field on null orbifolds can be found in
\cite{Berkooz:2005ym}.

 The paper is organized as follows. In Section $2$,
we introduce the fundamentals of our model while in Section $3$ we
derive the equations of motion for both the tachyon and the gauge
field dictated by a modified version of the DBI action for
unstable Dp branes. In Section $3$, having the equations of
motions at hand, we proceed by expressing the four dimensional
induced metric on the brane in terms of the constant of motions of
the system. The third section concludes, by setting forth the
complete formulas for the induced effective action on the probe
brane. Finally, in Section $4$, there is a discussion on the
effects of the tachyon- gauge boson system on the cosmological
expansion. We further analyze the effects of the tachyon on the
energy density for an exponentially decaying tachyon potential
trough a numerical integration of the equation of motions. Also,
useful conclusions are drawn especially in reference to former
works in this field.

%\newpage

\section{Geodesic Motion Of An Unstable D3-Brane }
In this section, we are going to unfold the basic features of our
proposed model while at the same time we will try to make contact
with what has been done in related investigations. We start with
the case of the gravitational background which is static in nature
and spherically symmetric. The metric that describes it is given
in the following general form
\begin{equation}
ds^2_{10}=g_{00}(r)dt^2+g(r)(d\overrightarrow{x})^2+
g_{rr}dr^2+g_{S}(r)d\Omega_{5} \label{1}
\end{equation}
where the timelike part of the metric $g_{00}$ is negative.

We also consider a rolling complex tachyon solution of an unstable
D$3$-brane minimally coupled to a massless gauge field
\begin{equation}
S_{3}=-T_{3}\int{d^4\xi}e^{-\phi}V(T)\sqrt{-det\tilde{K}_{\mu\nu}}-T_{3}\int{d^4\xi}\hat{C_{4}}\label{2}
\end{equation}
where
\begin{equation}
\tilde{K}_{\mu\nu}=K_{MN}\frac{\partial x^{M}}{\partial
x^{\mu}}\frac{\partial x^{N}}{\partial x^{\nu}} \label{3}
\end{equation}
and
\begin{equation}
K_{MN}=g_{MN}+2\pi{\alpha}'
F_{MN}+\frac{1}{2}(D_{M}TD_{N}T^{*}+c.c) \label{4}
\end{equation}
We ascribe capital indices M,N for the bulk coordinates while
$\mu,\nu$ are the one kept for the probe brane.

For completeness, we mention that $F_{\mu\nu}$ is the world volume
antisymmetric gauge field strength and $G_{MN}$ is the induced
metric on the D$3$-brane universe and $ \phi $ is the dilaton.
Note, that the RR field $\hat{C}(r)=C_{0...3}(r)$ is also included
in the action. In addition, the form of the complex tachyon field
and its covariant derivative are given as follows
\begin{equation}
D_{M}T=\partial_{M} T - iA_{M}T \label{5}
\end{equation}
\begin{equation}
T=T_{x}+iT_{y} \label{6}
\end{equation}

At this point, it is instructive to mention that our action which
describes the tachyon minimally coupled to a $U(1)$ gauge field
was initially proposed in a slightly different form in
\cite{Sen:2003tm} and implemented in a different context. In fact,
the analysis of \cite{Barnaby:2004dp,Cline:2005zy} was mainly
based upon investigating vortex like solutions when the tachyon
field is coupled to two different gauge fields one originating
from a D-brane and the other from a anti-brane. Therefore, the
gauge group of the system is of the form $ U(1)\times U(1)$. The
basic principles of this model were also adopted in later works
\cite{Barnaby:2004dp,Cline:2005zy} which were pertinent to
cosmological backgrounds viewed in terms of brane-antibrane
annihilation. In our action though, there is a single gauge field
and we also have considered that the transverse scalar fields of
the non-BPS brane are set to zero. Therefore, the tachyon
potential has no dependence on these scalars but only on the
tachyon field itself. Further, we demand that the tachyon is
restricted on the moving brane so it depends exclusively on time.

Given the fact that our system respects reparametrization
invariance, we are able to work on the static gauge, $
x^{\alpha}=\xi^{\alpha}, \alpha = 0,1,2,3 $. This particular
choice of gauge will really simplify the calculations. Also, due
to the angular part of the ambient space, the moving D$3$-brane
does inherit a non zero angular momentum in the transverse
dimensions. Having all of the above in mind the lagrangian is
given by

\begin{equation}
L=-\sqrt{T^2_{3}V^{2}(T)e^{-2\phi}[g^3(r)(|g_{00}|-g^{rr}\dot
r^2-g_{s}h_{ij}\dot \varphi^{i}\dot \varphi^{j}+\dot
T^2)-g^2(\alpha^{i}\dot A_{i}\dot T+\dot A_{i}\dot A^{i})]}+
\hat{C}(r) \label{7}
\end{equation}

In order to bring the lagrangian in this form, we performed the
following steps. First of all we considered without any loss of
generality the following ansatz for the gauge field

\begin{equation}
A^{i}(t)= A^{i}_{0}+ \tilde{A}^{i}(t) \label{8}
\end{equation}
where $A^{i}_{0}$ is independent of time

 One can also view $\tilde{A}^{i}(t)$ as very small
fluctuation of the gauge field around $A^{i}_{0}$. In other words,
we assume, that our massless gauge field is very weak. With this
assumption, only terms that are linear in the gauge field are
considered while higher order terms are small and thus neglected.
 In addition to that, we are interested in solutions where the
tachyon is really restrained to a region very close to the top of
the tachyon potential which in general is taken to be of the
following form

\begin{equation}
V(T)=\frac{V_{0}}{cosh(\frac{T}{T_{0}})} \label{9}
\end{equation}

At this point we note, that the potential which plays the role of
a varying tension has the proper asymptotic behavior, meaning that
$ V(T=0)=1 $ and $ V(T=\infty)=0 $. The significance of this
approximation will be apparent later on. However, in our case the
exact form of the tachyonic potential is not a crucial issue,
since later on we work in a region where the potential is flat.
Therefore under these two basic assumptions the form of
Eq.~(\ref{7}) comes is natural. Further, in the rest of this paper
the $ 2\pi\alpha' $ factor is suppressed and for convenience we
also drop the tilde from the gauge field. To further simplify our
calculations we use the $ A_{0}=0 $ gauge. In addition, the
imaginary part of the tachyon field is taken to be constant, i.e $
T_{y}= c $. Therefore, one can naturally expresses the constants
$\alpha^{i}$ as

\begin{equation}
\alpha^{i}=A^{i}_{0}c \label{10}
\end{equation}
where the index i runs over the three spatial dimensions.

With $ T_{3} $ we denote the tension of the probe D$3$-brane,
while the potential which drives the rolling tachyon is given by $
V(T) $. As with the tachyon, the three dimensional induced
components of the gauge field on the brane don't depend on any
other coordinates other than time.

We can also adopt the following redefinitions
$$ M=T^2_{3}V^2(T)e^{-2\phi}g^3(r)|g_{00}|$$
$$ B=T^2_{3}V^2(T)e^{-2\phi}g^{rr}(r)g^3(r)$$
$$ C=T^2_{3}V^2(T)e^{-2\phi}g_{s}(r)g^{3}(r)$$
$$ D=T^2_{3}V^2(T)e^{-2\phi}g^3(r)$$
$$ E=T^2_{3}V^2(T)e^{-2\phi}g^{2}(r)$$
Thus the lagrangian is recast in the compact form

\begin{equation}
L=-\sqrt{M-B\dot r^2-Ch_{ij}\dot \varphi^{i}\dot \varphi^{j}-D\dot
T^2-E(\alpha^{i}\dot A_{i}\dot T+ \dot A_{i}\dot A^{i})}+
\hat{C}(r) \label{11}
\end{equation}

\section{Classical Field Theory  Equations Of Motion}
In this section, since the general formalism of the paper is
already illustrated, we provide the classical equation of motion
for all fields in the Lagrangian through the variation principle.

The momenta are given by
\begin{equation} P_{r}=\frac{B\dot r}{\sqrt{M-B\dot
r^2-Ch_{ij}\dot \varphi^{i}\dot \varphi^{j}-D\dot
T^2-E(\alpha^{i}\dot A_{i}\dot T+ \dot A_{i}\dot A^{i})}}
\label{12}
\end{equation}

\begin{equation}
P_{i}=\frac{Ch_{ij}\dot \varphi^{j}}{\sqrt{M-B\dot r^2-Ch_{ij}\dot
\varphi^{i}\dot \varphi^{j}-D\dot T^2-E(\alpha^{i}\dot A_{i}\dot
T+ \dot A_{i}\dot A^{i})}} \label{13}
\end{equation}
%\newpage

In addition, the two dynamical fields (the tachyon and the gauge
boson) of the system are also endowed with the following momenta

\begin{equation}
P_{T}=\frac{E\alpha^{i}\dot A_{i}+2D\dot T}{2\sqrt{M-B\dot
r^2-Ch_{ij}\dot \varphi^{i}\dot \varphi^{j}-D\dot
T^2-E(\alpha^{i}\dot A_{i}\dot T+ \dot A_{i}\dot A^{i})}}
\label{14}
\end{equation}

\begin{equation}
P_{\dot A_{i}}=\frac{E(\alpha^{i}\dot T+2\dot
A^{i})}{2\sqrt{M-B\dot r^2-Ch_{ij}\dot \varphi^{i}\dot
\varphi^{j}-D\dot T^2-E(\alpha^{i}\dot A_{i}\dot T+ \dot A_{i}\dot
A^{i})}}\label{15}
\end{equation}

Given the fact that the Lagrangian doesn't exhibit any explicit
dependance upon time, it is expected, that the energy is
conserved. Therefore, the associated integral of motion is of the
form

\begin{equation}
H=\mathcal{E}=\frac{M}{\sqrt{M-B\dot r^2-Ch_{ij}\dot
\varphi^{i}\dot \varphi^{j}-D\dot T^2-E(\alpha^{i}\dot A_{i}\dot
T+ \dot A_{i}\dot A^{i})}}-\hat{C}(r) \label{16}
\end{equation}

We recall that since the ambient space is spherically symmetric,
angular momentum must be conserved, namely, $ h^{ij}\dot
\varphi^{i}\dot \varphi^{j}=l^{2}$( as it will be shown shortly
when examining closely the case of the near horizon limit of $
AdS_{5} \times S^{5} $ black hole background). Thus,

\begin{equation}
h_{ij}\dot \varphi^{i}\dot
\varphi^{j}=\frac{l^{2}}{C(C+l^2)}(M-B\dot r^2-Ch_{ij}\dot
\varphi^{i}\dot \varphi^{j}-D\dot T^2-E(\alpha^{i}\dot A_{i}\dot
T+ \dot A_{i}\dot A^{i})) \label{17}
\end{equation}

Next, we attempt to construct the equation of motion for the
tachyon field. To this end, we write

\begin{equation}
\frac{\partial L}{\partial \dot T}=\frac{E\alpha^{i}\dot
A_{i}+2D\dot T}{2\sqrt{M-B\dot r^2-Ch_{ij}\dot \varphi^{i}\dot
\varphi^{j}-D\dot T^2-E(\alpha^{i}\dot A_{i}\dot T+ \dot A_{i}\dot
A^{i})}} \label{18}
\end{equation}

\begin{equation}
\frac{\partial L}{\partial T}=-\frac{ dV}{
dT}\sqrt{T^2_{3}e^{-2\phi}[g^3(r)(|g_{00}|-g^{rr}\dot
r^2-g_{s}h_{ij}\dot \varphi^{i}\dot \varphi^{j}+\dot
T^2)-g^2(r)(\alpha^{i}\dot A_{i}\dot T+\dot A_{i}\dot A^{i})]}
\label{19}
\end{equation}

From Eq.~(\ref{19}) we conclude, that if one knows the actual form
of the potential which describes the motion of the rolling
tachyon, then solving this equation can be accomplished in
conjunction with the equation of motion of the gauge field. To
make things simpler we resort to working in a regime where the
tachyon rolls from the top of the potential where one can safely
take $ \frac{dV}{dT}\simeq 0 $ \cite{Jeong:2005py}.

%\newpage

Based upon this approximation one recovers the following integral
of motion

\begin{equation}
\tilde Q=\frac{E\alpha^{i}\dot A_{i}+2D\dot T}{2\sqrt{M-B\dot
r^2-Ch_{ij}\dot \varphi^{i}\dot \varphi^{j}-D\dot
T^2-E(\alpha^{i}\dot A_{i}\dot T+ \dot A_{i}\dot
A^{i})}}\label{20}
\end{equation}

As far as the the $ U(1) $ gauge field is concerned we set forth
the following integrals of motion

\begin{equation}
Q_{i}=\frac{E(2\dot A_{i}+\alpha_{i}\dot T)}{2\sqrt{M-B\dot
r^2-Ch_{ij}\dot \varphi^{i}\dot \varphi^{j}-D\dot
T^2-E(\alpha^{i}\dot A_{i}\dot T+ \dot A_{i}\dot
A^{i})}}\label{21}
\end{equation}

Henceforth, the RR term will not be explicitly written unless
stated otherwise. We proceed by providing the set of equations for
$(A_{i},T)$, the radial coordinate and the angular part of system
as functions of $( M, B, C, D, E, \alpha^{i} )$ and the constants
$ (l, \tilde{Q}, Q_{i}) $. After a laborious but straightforward
calculation one gets

\begin{equation}
\dot T^2=(\frac{M}{\mathcal{E}}\frac{4\tilde{Q}-2\alpha^{i}Q_{i}}
{4D-E\alpha^{i}\alpha_{i}})^2\label{22}
\end{equation}

\begin{equation}
(\dot A_{i})^2=(\frac{M}{2E\mathcal{E}}
[2Q_{i}-(\frac{4\tilde{Q}-2\alpha^{j}Q_{j}}
{4D-E\alpha^{j}\alpha_{j}})\alpha_{i}E])^2 \label{23}
\end{equation}

\begin{equation}
h_{ij}\dot \varphi^{i}\dot
\varphi^{j}=\frac{l^2}{C^2}\frac{M^2}{\mathcal{E}^2} \label{24}
\end{equation}

\begin{equation}
B\dot r^2=M-\frac{l^2}{C}\frac{M^2}{\mathcal{E}^2}
-\frac{M^2}{\mathcal{E}^2}-\frac{M^2}{E\mathcal{E}^2}Q^{i}Q_{i}
-\frac{M^2}{4\mathcal{E}^2}
\frac{(4\tilde{Q}-2\alpha^{i}Q_{i})^2}{4D-E\alpha^{i}\alpha_{i}}
\label{25}
\end{equation}

Last equation essentially tells us that the radial motion of the
brane is accompanied by the constraint dictated by the positivity
of its right hand side.

Before we conclude, it is worth mentioning that in the non
relativistic limit the whole analysis can be well approximated
through a more simplified version of Eq.~(\ref{11}). To be more
precise, in the low energy regime the DBI action can be expanded
in the following fashion

\begin{equation}
L=-\sqrt{M}-\frac{1}{2}\frac{B}{\sqrt{M}}\dot
r^2-\frac{1}{2}\frac{C}{\sqrt{M}}h_{ij}\dot \varphi^{i}\dot
\varphi^{j}-\frac{1}{2}\frac{D}{\sqrt{M}}\dot
T^2-\frac{1}{2}\frac{E}{\sqrt{M}}(\alpha^{i}\dot A_{i}\dot T+ \dot
A_{i}\dot A^{i})+\hat{C}(r) \label{26}
\end{equation}

While this equation can safely describe the evolution of the probe
universe in low energy scales (big scale factors), it fails to
capture the underlying physics at very early times where the
spatial dimensions are still very small. It is therefore very
essential for our analysis not to expand the square root in the
Lagrangian since we are mainly interested in calculating
quantities like the effective energy density and pressure on the
$D3$-brane as $a\rightarrow0$.

 A this point, the setup is ready to investigate the cosmological
implications of the moving probe, which is going to be described
in detail in the next section.

\section{Cosmological Evolution of the D3-brane Probe}

We are interested in giving a detailed account of the cosmological
implications of the current model. Basically, what is needed is
the exact form of the induced four dimensional metric on the
D$3$-brane. That amounts to

\begin{equation}
d\hat s^2=(g_{00}+g_{rr}\dot r^2+g_{S}h_{ij}\dot \varphi^{i}\dot
\varphi^{j})dt^2+d(\overrightarrow{x})^2 \label{27}
\end{equation}

%\newpage

Therefore, by direct substitution of equations Eq.~(\ref{24}) and
Eq.~(\ref{25}) into Eq.~(\ref{27}) the following form for the
metric is recovered

\begin{equation}
d\hat
s^2=-|g_{00}|^2(\frac{1}{\mathcal{E}^2}T^2_{3}V^2_{0}e^{-2\phi}g^3
+\frac{1}{\mathcal{E}^2}Q_{i}Q_{i}
+\frac{(2g\tilde{Q}-\alpha_{i}Q_{i})^2}{\mathcal{E}^2(4g^{2}-\alpha_{i}
\alpha_{i})})dt^2+g(d\overrightarrow{x})^2 \label{28}
\end{equation}

Close inspection reveals, that one can further simplify the above
metric in a more compact form

\begin{equation}
d\hat s^2=-d\eta^2+g(r(\eta))(d\overrightarrow{x})^2 \label{29}
\end{equation}

while the cosmic time is defined as

\begin{equation}
d\eta=\frac{|g_{00}|}{\mathcal{E}}\sqrt{T^2_{3}V^2_{0}e^{-2\phi}g^3
+Q_{i}Q_{i}
+\frac{(2g\tilde{Q}-\alpha_{i}Q_{i})^2}{(4g^{2}-\alpha_{i}
\alpha_{i})}}dt \label{30}
\end{equation}

 For completeness, we mention, that one can compute the apparent Ricci scalar
 curvature of the four dimensional D$3$ brane universe in terms of
 the induced effective energy density as follows
\begin{equation}
 R_{4-d}=8\pi G(\rho_{eff}-3p_{eff})=8\pi
 G(4+a\frac{\partial}{\partial a})\rho_{eff} \label{31}
 \end{equation}

Apparently, one infers, that the argument of the radical in
Eq.~(\ref{30}) is not only non-positive  but it is also singular
as well when $ 4g^2(r)=\alpha_{i}\alpha_{i}$ unlike in models like
\cite{Kehagias:1999vr}. However, this is not that surprising
because of the nontrivial nature of the gauge field-tachyon
interaction. Nonetheless, we are mainly interested in the regime
where the scale factor of the brane goes to zero. At that limit it
is easy to check, that the argument in the radical is strictly
nonnegative (for nonzero values of the $Q_{i}$ charges) while the
singularity has been avoided.

We can also choose the scale factor of the universe to obey the
relation $ a^2=g$, as it is widely used in mirage cosmology.
Therefore, the four-dimensional Friedman equation can be brought
in the form

\begin{equation}
(\frac{\dot{a}}{a})^2=
\frac{\mathcal{E}^2-(l^2g_{S}^{-1}+R^2)|g_{00}|}{4g_{rr}|g_{00}|R^2}
(\frac{g'}{g})^2 \label{32}
\end{equation}

where

\begin{equation}
R^2=T^2_{3}V^2_{0}e^{-2\phi}g^3 +Q_{i}Q_{i}
+\frac{(2g\tilde{Q}-\alpha_{i}Q_{i})^2}{(4g^{2}-\alpha_{i}
\alpha_{i})} \label{33}
\end{equation}

The dot stands for differentiation with respect to cosmic time
$\eta$, while the prime stands for derivatives with respect to the
radial coordinate r. Also, the Hubble constant is defined as
$H=\dot a /a$. On the other hand, Eq.~(\ref{31}) can also be
viewed in terms of the effective matter density on the probe brane

\begin{equation}
\frac{8\pi G}{3}\rho_{eff}=
\frac{\mathcal{E}^2-(l^2g_{S}^{-1}+R^2)|g_{00}|}{4g_{rr}|g_{00}|R^2}
(\frac{g'}{g})^2 \label{34}
\end{equation}

Interestingly enough, an observer residing on the moving
D$3$-brane feels the cosmological evolution of the universe as a
continuous change of the scale factor $a(\eta)$ which depends on
the radial position r.

As an example, one can study the very characteristic example of
the near-horizon limit of a black hole in an
$AdS_{5}\times{S^{5}}$ spacetime. The linear element of this
ambient space is

\begin{equation}
ds^2=\frac{r^2}{L^2}(-f(r)dt^2+(d\overrightarrow{x}^2))
+\frac{L^2}{r^2}\frac{dr^2}{f(r)}+L^2d\Omega^2_{5} \label{35}
\end{equation}

where $f(r)=1-\frac{r_{0}^4}{r^4}$ and $\hat{C}(r)=
\frac{r^4}{L^4}-\frac{r_{0}^4}{2L^4}$

In addition, the following identifications are made

$$g_{00}(r)=-\frac{r^2}{L^2}f(r)$$
$$g(r)=\frac{r^2}{L^2}$$
$$g_{rr}(r)=\frac{L^2}{r^2}\frac{1}{f(r)}$$
$$g_{S}(r)=L^2$$

which lead to the analogue of Friedman equation on the probe brane

\begin{equation}
H^2=\frac{8\pi G}{3}\rho_{eff}=
\frac{1}{a^2L^2F^2}((\mathcal{E}+a^4)^2-a^2(1-\frac{r_{0}^4}{L^4a^4})
(F^2+\frac{l_{S}^2}{L^2})) \label{36}
\end{equation}
The constant term accompanying the RR term is absorbed in
$\mathcal{E}$. The explicit form of F is given as a function of
the scale factor of the universe in the following form

\begin{equation}
F^2=T^2_{3}V_{0}^2e^{-2\phi}a^6+Q_{i}Q_{i}
+\frac{(2\tilde{Q}a^2-\alpha_{i}Q_{i})^2}{4a^4-\alpha_i\alpha_{i}}
\label{37}
\end{equation}

Having laid out the basic constituents of our model, it is time to
make some remarks with regarding its effects on cosmological
evolution. Our results are in agreement with the vanishing gauge
field case \cite{Jeong:2005py} in which the effective density on
the brane is given in the following form

\begin{equation}
\frac{8\pi G}{3}\rho_{eff}=
\frac{1}{a^2(a^6e^{-2\phi}V_{0}^2T^2_{3}+Q^2)L^2}
[E^2-(1-\frac{r_{0}^4}{L^4}\frac{1}
{a^4})a^2(a^6e^{-2\phi}V_{0}^2+Q^2+\frac{l^2}{L^2})] \label{38}
\end{equation}
%\newpage

After performing the following substitutions where,
$Q\rightarrow{\tilde{Q}}, E\rightarrow{\mathcal{E}}$, then it
becomes apparent that equations Eq.~(\ref{36}) and Eq.~(\ref{38})
are identical in the case of vanishing Ramond-Ramond field. As it
was also pointed out in \cite{Jeong:2005py} the presence of the
tachyonic charge $\tilde{Q}$, renders the effective brane density
much less divergent than in \cite{Kehagias:1999vr}, where
$\rho_{eff}\sim a^{-8}$ as the scale factor approaches zero. One
may also consider the case $l=r_{0}=0$. Here, the universe first
expands when at some point in its history it stops and finally
recollapses.

In a different limit where the tachyon field vanishes and the
probe brane is stable ($V_{0}=1, \alpha_{i}=0, \tilde{Q}=0$), we
were also able to reproduce equation (4.7) of
\cite{Kehagias:1999vr} after proper redefinition of variables.
Another possible model is the one where both the gauge field and
the tachyon are decoupled to each other ($\alpha_{i}=0$).
Additionally, the positivity of the right hand side of
Eq.~(\ref{37}) is restored and in general the complete form of the
energy density reads as follows

\begin{equation}
\frac{8\pi
G}{3}\rho_{eff}=\frac{1}{a^2(a^6e^{-2\phi}V_{0}^2T^2_{3}+\tilde{Q}^2+Q_{i}^2)L^2}
[(\mathcal{E}+a^4)^2-(1-\frac{r_{0}^4}{L^4}\frac{1}
{a^4})a^2(a^6e^{-2\phi}V_{0}^2+\tilde{Q}^2+Q_{i}^2+\frac{l^2}{L^2})]
\label{39}
\end{equation}

where $Q_{i}^2=Q_{i}Q_{i}=|\overrightarrow{Q}|^2$.

Of course the most general model derived by our approach is the
one dictated by Eq.~(\ref{35}). In particular, in the limit where
$ a\rightarrow 0$ one gets

\begin{equation}
F^2=T_{3}^2V_{0}e^{-2\phi}a^6+|\overrightarrow{Q}|^2sin^2{\theta}
\label{40}
\end{equation}

where $\theta$ denotes the angle between vectors $\alpha_{i}$ and
$Q_{i}$. In other words, the effective density is insensitive to
the magnitude of $\alpha_{i}$ vector which in turn is proportional
to the vacuum expectation value $A^{i}_{0}$ of the gauge boson
Eq.~(\ref{10}). Note, that the singular behavior that the $F^2$
term exhibited in the most general case, has disappeared at the
very early stages of the evolution of the universe. Further,
equation Eq.~(\ref{38}) shows that there are sectors of the theory
where $ \rho_{eff}\sim a^{-8}$ when the cofactor of
$|\overrightarrow{Q}|^2$  term is zero. To put it simply, in the
case where a minimal coupling is included the effective density on
the probe may have similar form with the one of a stable D3-brane
universe. In all other cases this novel effect doesn't occur due
to the non vanishing charges. It is also apparent, that in
Eq.~(\ref{36}) there is no explicit dependance on the tachyon
charge and only the gauge field charges survives. This could be
easily attributed to the well known fact that radiation dominates
matter at early times in an expanding universe.

More precisely, in order that we acquire a more detailed picture
of the properties of the cosmic fluid the exact form of the
pressure needs to determined. The formula that associates the
effective pressure with the density can be expressed in terms of
the scale factor as follows
\begin{equation}
p_{eff}=-\rho_{eff}-\frac{1}{3}a\frac{\partial\rho_{eff}}{\partial{a}}
\label{41}
\end{equation}

The last equation can be obtained through a direct implementation
of Raychaudhuri equation which states as follows

\begin{equation}
\frac{\ddot{a}}{a}=(1+ \frac{1}{2}a\frac{\partial}{\partial
a})\frac{8 \pi G}{3}\rho_{eff}=-\frac{4 \pi
G}{3}(\rho_{eff}+3p_{eff}) \label{42}
\end{equation}

At the limit where the scale factor goes to zero the pressure
reads

\begin{eqnarray}
p_{eff}= \frac{1}{8\pi GL^2}\frac{r_{0}^4}{L^4}
 (1+\frac{l^2}{L^2}\frac{1}{Q_{i}^2sin^2{\theta}})\frac{1}{a^4}\nonumber\\
+(\frac{\mathcal{E}^2}{8\pi
 GL^2})(-\frac{1}{(Q_{i}^2sin^2{\theta})^2}
+\frac{8r_{0}^4}{L^4}\frac{l^2}{L^2}\frac{\alpha_{i}Q_{i}\tilde{Q}}
{\alpha_{i}^2(Q_{i}^2sin^2{\theta})^2})\frac{1}{a^2} \nonumber \\
+\frac{3}{8\pi
 GL^2}(1+\frac{l^2}{L^2}\frac{1}{Q_{i}^2sin^2{\theta}}
+\frac{8\mathcal{E}^2}{3}
\frac{\alpha_{i}Q_{i}\tilde{Q}}{\alpha_{i}^2(Q_{i}^2sin^2{\theta})^2}
-\frac{16}{3}\frac{r_{0}^4}{L^4}\frac{l^2}{L^2}\frac{(\alpha_{i}Q_{i})^2}{\alpha_{i}^4})
\label{43}
\end{eqnarray}

while the effective density is of the form

\begin{equation}
\rho_{eff}=\frac{3}{8\pi
GL^2}\frac{r_{0}}{L^4}(1+\frac{l^2}{L^2}\frac{1}{Q_{i}sin^2{\theta}})\frac{1}{a^4}
+\frac{3\mathcal{E}^2}{8\pi
GL^2}\frac{1}{(Q_{i}^2sin^2{\theta})^2}\frac{1}{a^2}-
\frac{3}{8\pi
GL^2}(1+\frac{l^2}{L^2}\frac{1}{Q_{i}^2sin^2{\theta}}) \label{44}
\end{equation}

For an  ideal fluid the state equation reads $ p_{eff}=w\rho_{eff}
$ so  the constant $w$ encapsulates all the details of its
structure. To be more precise, in the case where  minimal coupling
is included $w$ takes the following values
$w=\frac{1}{3},-\frac{1}{3},-1$. Initially, the moving brane is
radiation dominated but at later times it has the state equation
of a universe filled with a cosmic fluid that resembles a nonzero
cosmological constant. Thus, even though $w=-1$ the corresponding
energy density is negative which is indicative of the unstable
nature of the tachyon. For completeness we note that causality is
respected at all times since the constraint $|w|\leq{1}$ is
satisfied. The above values for $w$ are reminiscent of
quintessence in which the accelerated expansion of the universe
occurs when $-1<w<-\frac{1}{3}$. So for the probe brane, $w$ falls
just outside of those bounds. A more extensive analysis on those
limits with regard to the interplay between string theory and
quintessential cosmological models can be found at
\cite{Hellerman:2001yi,Fischler:2001yj}.

Finally one finds, that the term proportional to $a^{-2}$ in the
energy density exhibits the same behavior as a negative curvature
term which accelerates the expansion of the universe. In the most
general case where the minimal coupling is taken into account, $w$
can take all shorts of values, however as long as we are looking
back into the very early history of the expanding probe brane we
observe that $p_{eff}=\frac{1}{3}\rho_{eff}$ holds. In addition,
it is obvious that the inclusion of the minimal coupling will
certainly give rise to regions where causality is violated,
depending on the actual value of the free parameters of the theory
which affect the state equation of the mirage mater. There is no
doubt therefore, that the presence of the nontrivial coupling
between the tachyon and the gauge field in the Lagrangian cannot
drastically alter the state equation of the brane universe when
the scale factor remains very small.

In has become more or less clear that our analysis focuses
primarily on the effects of the kinetic term of the tachyon on the
cosmological expansion rather than on it's potential. However, a
more detailed picture can be obtained by including the explicit
form of the potential of the rolling tachyon. Even though it is
really tough to get an analytic solution for the energy density we
can still extract some useful information regarding the physics of
our system through numerical analysis. To this end as well as to
simplify things, we consider just the case where the tachyon lives
on the brane in the absence of any fluxes. We further assume that
the angular momentum $l$ is zero. By  utilizing
Eq.~(\ref{13},\ref{14},\ref{16},\ref{18},\ref{19}) which involve
the equations of motions for the tachyon, the radial motion and
the conservation of energy, one observes that the energy density
is much less divergent as the scale factor goes to zero when
tachyonic degrees of freedom are present on the brane. This is
well transpired in our graph for a given set of initial conditions
(see fig.1), while we also choose $T_3e^{-\phi}=1, \mathcal{E}=1,
L=1$. In addition, the type of tachyon potential implemented in
our analysis is exponentially decaying $V(T)\sim{e^{-T}}$
\cite{Sen:2002nu}. Therefore, for appropriate initial conditions
we reach the same conclusions as in the case where the tachyon is
close to the top of the potential.

\begin{figure}
\includegraphics{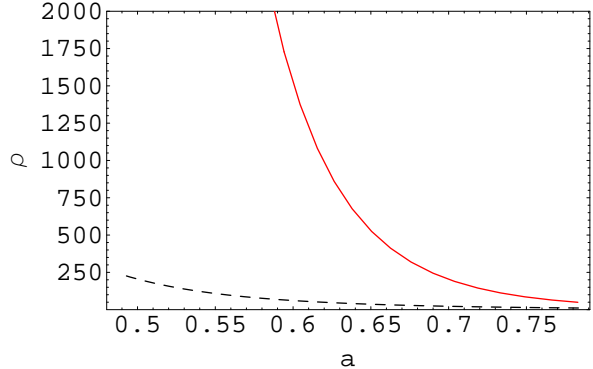}
\caption{\label{fig1}Plot of the effective brane density
$\rho_{eff}$ as a function of the scale factor of the universe
$a$. The red solid line represents the density when the tachyon is
on the brane while the dark dashed line at the bottom shows the
density when the tachyon is absent (stable D$3$-brane). The
initial conditions used for both cases are $r=2,
\frac{\partial{r}}{\partial{t}}=0.2, T=0,
\frac{\partial{T}}{\partial{t}}=1.0$, regarding the radial and
tachyonic equations of motions respectively at $t=0.1$.}
\end{figure}

\section{Conclusions}

In this paper we have extended the mirage cosmological models to
the case of tachyon matter living on the moving D$3$-brane which
is coupled to $U(1)$ gauge field. Under the approximations that
the gauge field is very weak and that the complex tachyon field is
restrained on the top of the potential we derived the explicit
form of the equations of motion for both dynamical fields. In
fact, due to both the spherical symmetry of the $S^5$ part of the
ambient space and energy conservation, the system is rendered
integrable. Expressing the four dimensional induced metric in
terms of the constants of motion enabled as to recast the metric
in a form that describes a flat four dimensional universe.
Eventually, the effective brane density was obtained giving rise
to several cosmological interpretations. For instance, it was
shown that the presence of the tachyon makes the cosmological
expansion much less divergent at the early times of the evolution
of the universe as it was pointed out in former works. The
coupling between the two fields doesn't alter this basic feature
leading one to the conclusion that tachyon field acts like a
regulator of the cosmological expansion in most cases. However, we
did actually show, that in some regions of the theory the
regulatory effect of the tachyon on the effective brane density is
absent since it looks that the tachyon is unable to soften the
degree of divergence that the effective energy density exhibits
when the size of the universe is very small. Of course, in a more
general treatment (which is out of the scope of this paper) where
the gauge field becomes strong, one may lead to a different
conclusion. Finally through a numerical analysis we show that even
if one studies the motion of the tachyonic brane in the absence of
gauge degrees of freedom the energy density is still softened for
small scale factors given a decaying like tachyon potential. We
hope, that future developments in this area will emerge, so that
new aspects of tachyon cosmology will shed more light on the
mechanisms affecting the very early stages of
the universe.\\

\noindent
{\bf Acknowledgements}\\

%\begin{acknowlegments}
The author would like to express his gratitude to Ioannis Bakas
for his constant motivation, inspiration, support and for his
comments on the manuscript. I would like to thank Elias Kiritsis
for correspondence and suggestions, Pran Nath, Tomasz Taylor and
Yogi Srivastava, for reading the paper and for their comments.
Also, I thank Allan Widom for very useful and stimulating
discussions and suggestions during the most crucial stages of this
work. This work was supported in part by U.S. National Science
Foundation grant NSF-PHY-$0546568$.

%\end{acknowlegments}


\begin{thebibliography}{98}
%\cite{Guth:1980zm}
\bibitem{Guth:1980zm}
  A.~H.~Guth,
  ``The Inflationary Universe: A Possible Solution To The Horizon And Flatness
  Problems,''
  Phys.\ Rev.\ D {\bf 23}, 347 (1981).
  %%CITATION = PHRVA,D23,347;%%

%Phys.\ Rev.\ D {\bf 23}, 347 (1981).
%A. Guth, Phys. Rev.{\bf D23 },347 (1981).


%\cite{Alexander:2000xv}
\bibitem{Alexander:2000xv}
  S.~Alexander, R.~H.~Brandenberger and D.~Easson,
  ``Brane gases in the early universe,''
  Phys.\ Rev.\ D {\bf 62}, 103509 (2000)
  [arXiv:hep-th/0005212].
  %%CITATION = HEP-TH 0005212;%%

%\cite{Brandenberger:2001kj}
\bibitem{Brandenberger:2001kj}
  R.~Brandenberger, D.~A.~Easson and D.~Kimberly,
  ``Loitering phase in brane gas cosmology,''
  Nucl.\ Phys.\ B {\bf 623}, 421 (2002)
  [arXiv:hep-th/0109165].
  %%CITATION = HEP-TH 0109165;%%

%\cite{Brandenberger:1988aj}
\bibitem{Brandenberger:1988aj}
  R.~H.~Brandenberger and C.~Vafa,
  ``Superstrings In The Early Universe,''
  Nucl.\ Phys.\ B {\bf 316}, 391 (1989).
  %%CITATION = NUPHA,B316,391;%%

%\cite{Tseytlin:1991xk}
\bibitem{Tseytlin:1991xk}
  A.~A.~Tseytlin and C.~Vafa,
  ``Elements of string cosmology,''
  Nucl.\ Phys.\ B {\bf 372}, 443 (1992)
  [arXiv:hep-th/9109048].
  %%CITATION = HEP-TH 9109048;%%


%\cite{Garousi:2000tr}
\bibitem{Garousi:2000tr}
  M.~R.~Garousi,
  %``Tachyon couplings on non-BPS D-branes and Dirac-Born-Infeld action,''
  Nucl.\ Phys.\ B {\bf 584}, 284 (2000)
  [arXiv:hep-th/0003122].
  %%CITATION = HEP-TH 0003122;%%


%\cite{Bergshoeff:2000dq}
%\bibitem{Bergshoeff:2000dq}
  E.~A.~Bergshoeff, M.~de Roo, T.~C.~de Wit, E.~Eyras and S.~Panda,
  %``T-duality and actions for non-BPS D-branes,''
  JHEP {\bf 0005}, 009 (2000)
  [arXiv:hep-th/0003221].
  %%CITATION = HEP-TH 0003221;%%


  %\cite{Mazumdar:2001mm}
%\bibitem{Mazumdar:2001mm}
  A.~Mazumdar, S.~Panda and A.~Perez-Lorenzana,
  %``Assisted inflation via tachyon condensation,''
  Nucl.\ Phys.\ B {\bf 614}, 101 (2001)
  [arXiv:hep-ph/0107058].
  %%CITATION = HEP-PH 0107058;%%

%\cite{Kluson:2000iy}
%\bibitem{Kluson:2000iy}
J.~Kluson,
%``Proposal for non-BPS D-brane action,''
Phys.\ Rev.\ D {\bf 62}, 126003 (2000) [arXiv:hep-th/0004106].
%%CITATION = HEP-TH 0004106;%%





%\cite{Choudhury:2002xu}
%\bibitem{Choudhury:2002xu}
  D.~Choudhury, D.~Ghoshal, D.~P.~Jatkar and S.~Panda,
  %``On the cosmological relevance of the tachyon,''
  Phys.\ Lett.\ B {\bf 544}, 231 (2002)
  [arXiv:hep-th/0204204].
  %%CITATION = HEP-TH 0204204;%%


%\cite{Sen:2002nu}
\bibitem{Sen:2002nu}
  A.~Sen,
  ``Rolling tachyon,''
  JHEP {\bf 0204}, 048 (2002)
  [arXiv:hep-th/0203211].
  %%CITATION = HEP-TH 0203211;%%

%\cite{Sen:2002in}
%\bibitem{Sen:2002in}
  A.~Sen,
  ``Tachyon matter,''
  JHEP {\bf 0207}, 065 (2002)
  [arXiv:hep-th/0203265].
  %%CITATION = HEP-TH 0203265;%%

%\cite{Sen:2002an}
%\bibitem{Sen:2002an}
  A.~Sen,
  ``Field theory of tachyon matter,''
  Mod.\ Phys.\ Lett.\ A {\bf 17}, 1797 (2002)
  [arXiv:hep-th/0204143].
  %%CITATION = HEP-TH 0204143;%%


%\cite{Kutasov:2000qp}
\bibitem{Kutasov:2000qp}
  D.~Kutasov, M.~Marino and G.~W.~Moore,
  ``Some exact results on tachyon condensation in string field theory,''
  JHEP {\bf 0010}, 045 (2000)
  [arXiv:hep-th/0009148].
  %%CITATION = HEP-TH 0009148;%%

%\cite{Kutasov:2004dj}
\bibitem{Kutasov:2004dj}
  D.~Kutasov,
  ``D-brane dynamics near NS5-branes,''
  arXiv:hep-th/0405058.
  %%CITATION = HEP-TH 0405058;%%


%\cite{Frolov:2002rr}
\bibitem{Frolov:2002rr}
  A.~V.~Frolov, L.~Kofman and A.~A.~Starobinsky,
  ``Prospects and problems of tachyon matter cosmology,''
  Phys.\ Lett.\ B {\bf 545}, 8 (2002)
  [arXiv:hep-th/0204187].
  %%CITATION = HEP-TH 0204187;%%


%\cite{Kofman:2002rh}
\bibitem{Kofman:2002rh}
  L.~Kofman and A.~Linde,
  ``Problems with tachyon inflation,''
  JHEP {\bf 0207}, 004 (2002)
  [arXiv:hep-th/0205121].
  %%CITATION = HEP-TH 0205121;%%


%\cite{Felder:2002sv}
\bibitem{Felder:2002sv}
  G.~N.~Felder, L.~Kofman and A.~Starobinsky,
  ``Caustics in tachyon matter and other Born-Infeld scalars,''
  JHEP {\bf 0209}, 026 (2002)
  [arXiv:hep-th/0208019].
  %%CITATION = HEP-TH 0208019;%%

%\cite{Shiu:2002qe}
\bibitem{Shiu:2002qe}
  G.~Shiu and I.~Wasserman,
  ``Cosmological constraints on tachyon matter,''
  Phys.\ Lett.\ B {\bf 541}, 6 (2002)
  [arXiv:hep-th/0205003].
  %%CITATION = HEP-TH 0205003;%%


%\cite{Gibbons:2002md}
\bibitem{Gibbons:2002md}
  G.~W.~Gibbons,
  ``Cosmological evolution of the rolling tachyon,''
  Phys.\ Lett.\ B {\bf 537}, 1 (2002)
  [arXiv:hep-th/0204008].
  %%CITATION = HEP-TH 0204008;%%

%\cite{Feinstein:2002aj}
\bibitem{Feinstein:2002aj}
  A.~Feinstein,
  ``Power-law inflation from the rolling tachyon,''
  Phys.\ Rev.\ D {\bf 66}, 063511 (2002)
  [arXiv:hep-th/0204140].
  %%CITATION = HEP-TH 0204140;%%

%\cite{Padmanabhan:2002sh}
\bibitem{Padmanabhan:2002sh}
  T.~Padmanabhan and T.~R.~Choudhury,
  ``Can the clustered dark matter and the smooth dark energy arise from the
  same scalar field?,''
  Phys.\ Rev.\ D {\bf 66}, 081301 (2002)
  [arXiv:hep-th/0205055].
  %%CITATION = HEP-TH 0205055;%%

%\cite{Kim:2002zr}
\bibitem{Kim:2002zr}
  C.~j.~Kim, H.~B.~Kim and Y.~b.~Kim,
  ``Rolling tachyons in string cosmology,''
  Phys.\ Lett.\ B {\bf 552}, 111 (2003)
  [arXiv:hep-th/0210101].
  %%CITATION = HEP-TH 0210101;%%

%\cite{Shiu:2002xp}
\bibitem{Shiu:2002xp}
  G.~Shiu, S.~H.~H.~Tye and I.~Wasserman,
  ``Rolling tachyon in brane world cosmology from superstring field theory,''
  Phys.\ Rev.\ D {\bf 67}, 083517 (2003)
  [arXiv:hep-th/0207119].
  %%CITATION = HEP-TH 0207119;%%

%\cite{Gorini:2004by}
\bibitem{Gorini:2004by}
  V.~Gorini, A.~Kamenshchik, U.~Moschella and V.~Pasquier,
  ``The Chaplygin gas as a model for dark energy,''
  arXiv:gr-qc/0403062.
  %%CITATION = GR-QC 0403062;%%


%\cite{Piao:2003sc}
\bibitem{Piao:2003sc}
  Y.~S.~Piao and Y.~Z.~Zhang,
  ``Thermal tachyon inflation,''
  arXiv:hep-th/0307074.
  %%CITATION = HEP-TH 0307074;%%

%\cite{Panigrahi:2004qr}
\bibitem{Panigrahi:2004qr}
  K.~L.~Panigrahi,
  %``D-brane dynamics in Dp-brane background,''
  Phys.\ Lett.\ B {\bf 601}, 64 (2004)
  [arXiv:hep-th/0407134].
  %%CITATION = HEP-TH 0407134;%%


%\cite{Gibbons:2003gb}
\bibitem{Gibbons:2003gb}
  G.~W.~Gibbons,
  ``Thoughts on tachyon cosmology,''
  Class.\ Quant.\ Grav.\  {\bf 20}, S321 (2003)
  [arXiv:hep-th/0301117].
  %%CITATION = HEP-TH 0301117;%%


%\cite{Gibbons:1984kp}
\bibitem{Gibbons:1984kp}
  G.~W.~Gibbons,
  ``Aspects Of Supergravity Theories,''
Print-85-0061 (CAMBRIDGE)
\href{http://www.slac.stanford.edu/spires/find/hep/www?r=print-85-0061\%2F(cambridge)}{SPIRES
entry} {\it Three lectures given at GIFT Seminar on Theoretical
Physics, San Feliu de Guixols, Spain, Jun 4-11, 1984}


%\cite{Maldacena:2000yy}
\bibitem{Maldacena:2000yy}
  J.~M.~Maldacena and C.~Nunez,
  ``Towards the large N limit of pure N = 1 super Yang Mills,''
  Phys.\ Rev.\ Lett.\  {\bf 86}, 588 (2001)
  [arXiv:hep-th/0008001].
  %%CITATION = HEP-TH 0008001;%%



%\cite{Ohta:2004wk}
%\bibitem{Ohta:2004wk}
  N.~Ohta,
  ``Accelerating cosmologies and inflation from M / superstring theories,''
  Int.\ J.\ Mod.\ Phys.\ A {\bf 20}, 1 (2005)
  [arXiv:hep-th/0411230].
  %%CITATION = HEP-TH 0411230;%%

%\cite{Neupane:2005nb}
%\bibitem{Neupane:2005nb}
  I.~P.~Neupane and D.~L.~Wiltshire,
  %``Cosmic acceleration from M theory on twisted spaces,''
  arXiv:hep-th/0504135.
  %%CITATION = HEP-TH 0504135;%%


%\cite{Randall:1999ee}
\bibitem{Randall:1999ee}
  L.~Randall and R.~Sundrum,
  ``A large mass hierarchy from a small extra dimension,''
  Phys.\ Rev.\ Lett.\  {\bf 83}, 3370 (1999)
  [arXiv:hep-ph/9905221].
  %%CITATION = HEP-PH 9905221;%%


%\cite{Randall:1999vf}
\bibitem{Randall:1999vf}
  L.~Randall and R.~Sundrum,
  ``An alternative to compactification,''
  Phys.\ Rev.\ Lett.\  {\bf 83}, 4690 (1999)
  [arXiv:hep-th/9906064].
  %%CITATION = HEP-TH 9906064;%%


%\cite{Horava:1996ma}
\bibitem{Horava:1996ma}
  P.~Horava and E.~Witten,
  ``Eleven-Dimensional Supergravity on a Manifold with Boundary,''
  Nucl.\ Phys.\ B {\bf 475}, 94 (1996)
  [arXiv:hep-th/9603142].
  %%CITATION = HEP-TH 9603142;%%



%\cite{Arkani-Hamed:1998rs}
\bibitem{Arkani-Hamed:1998rs}
  N.~Arkani-Hamed, S.~Dimopoulos and G.~R.~Dvali,
  ``The hierarchy problem and new dimensions at a millimeter,''
  Phys.\ Lett.\ B {\bf 429}, 263 (1998)
  [arXiv:hep-ph/9803315].
  %%CITATION = HEP-PH 9803315;%%


%\cite{Antoniadis:1998ig}
\bibitem{Antoniadis:1998ig}
  I.~Antoniadis, N.~Arkani-Hamed, S.~Dimopoulos and G.~R.~Dvali,
  ``New dimensions at a millimeter to a Fermi and superstrings at a TeV,''
  Phys.\ Lett.\ B {\bf 436}, 257 (1998)
  [arXiv:hep-ph/9804398].
  %%CITATION = HEP-PH 9804398;%%


%\cite{Cremades:2002dh}
\bibitem{Cremades:2002dh}
  D.~Cremades, L.~E.~Ibanez and F.~Marchesano,
  %``Standard model at intersecting D5-branes: Lowering the string scale,''
  Nucl.\ Phys.\ B {\bf 643}, 93 (2002)
  [arXiv:hep-th/0205074].
  %%CITATION = HEP-TH 0205074;%%


%\cite{Kokorelis:2002qi}
%\bibitem{Kokorelis:2002qi}
  C.~Kokorelis,
  %``Exact standard model structures from intersecting D5-branes,''
  Nucl.\ Phys.\ B {\bf 677}, 115 (2004)
  [arXiv:hep-th/0207234].
  %%CITATION = HEP-TH 0207234;%%


%\cite{Maldacena:1997re}
\bibitem{Maldacena:1997re}
  J.~M.~Maldacena,
  ``The large N limit of superconformal field theories and supergravity,''
  Adv.\ Theor.\ Math.\ Phys.\  {\bf 2}, 231 (1998)
  [Int.\ J.\ Theor.\ Phys.\  {\bf 38}, 1113 (1999)]
  [arXiv:hep-th/9711200].
  %%CITATION = HEP-TH 9711200;%%

%\cite{Kehagias:1999vr}
\bibitem{Kehagias:1999vr}
  A.~Kehagias and E.~Kiritsis,
  ``Mirage cosmology,''
  JHEP {\bf 9911}, 022 (1999)
  [arXiv:hep-th/9910174].
  %%CITATION = HEP-TH 9910174;%%


%\cite{Papantonopoulos:2000xs}
\bibitem{Papantonopoulos:2000xs}
  E.~Papantonopoulos,
  ``Brane inflation from mirage cosmology,''
  arXiv:hep-th/0011051.
  %%CITATION = HEP-TH 0011051;%%



%\cite{Youm:2000mi}
\bibitem{Youm:2000mi}
  D.~Youm,
  ``Brane inflation in the background of D-brane with NS B field,''
  Phys.\ Rev.\ D {\bf 63}, 125019 (2001)
  [arXiv:hep-th/0011024].
  %%CITATION = HEP-TH 0011024;%%


%\cite{Youm:2000ke}
\bibitem{Youm:2000ke}
  D.~Youm,
  ``Closed universe in mirage cosmology,''
  Phys.\ Rev.\ D {\bf 63}, 085010 (2001)
  [Erratum-ibid.\ D {\bf 63}, 129902 (2001)]
  [arXiv:hep-th/0011290].
  %%CITATION = HEP-TH 0011290;%%


%\cite{Brax:2002qw}
\bibitem{Brax:2002qw}
  P.~Brax and D.~A.~Steer,
  ``Non-BPS brane cosmology,''
  JHEP {\bf 0205}, 016 (2002)
  [arXiv:hep-th/0204120].
  %%CITATION = HEP-TH 0204120;%%




%\cite{Bozhilov:2002fa}
\bibitem{Bozhilov:2002fa}
  P.~Bozhilov,
  ``Probe branes dynamics in nonconstant background fields,''
  arXiv:hep-th/0206253.
  %%CITATION = HEP-TH 0206253;%%



%\cite{Kim:2002hk}
\bibitem{Kim:2002hk}
  J.~Y.~Kim,
  ``Mirage cosmology in M-theory,''
  Phys.\ Lett.\ B {\bf 548}, 1 (2002)
  [arXiv:hep-th/0203084].
  %%CITATION = HEP-TH 0203084;%%


%\cite{Piao:2002vf}
\bibitem{Piao:2002vf}
  Y.~S.~Piao, R.~G.~Cai, X.~m.~Zhang and Y.~Z.~Zhang,
  %``Assisted tachyonic inflation,''
  Phys.\ Rev.\ D {\bf 66}, 121301 (2002)
  [arXiv:hep-ph/0207143].
  %%CITATION = HEP-PH 0207143;%%




%\cite{Boehm:2002kf}
\bibitem{Boehm:2002kf}
  T.~Boehm and D.~A.~Steer,
  ``Perturbations on a moving D3-brane and mirage cosmology,''
  Phys.\ Rev.\ D {\bf 66}, 063510 (2002)
  [arXiv:hep-th/0206147].
  %%CITATION = HEP-TH 0206147;%%



%\cite{Papantonopoulos:2004au}
\bibitem{Papantonopoulos:2004au}
  E.~Papantonopoulos and V.~Zamarias,
  ``AdS/CFT correspondence and the reheating of the brane-universe,''
  JHEP {\bf 0410}, 051 (2004)
  [arXiv:hep-th/0408227].
  %%CITATION = HEP-TH 0408227;%%

%\cite{Iwashita:2005wi}
\bibitem{Iwashita:2005wi}
  Y.~Iwashita, T.~Shiromizu, K.~Takahashi and S.~Fujii,
  %``Gravity is controlled by cosmological constant,''
  Phys.\ Rev.\ D {\bf 71}, 083518 (2005)
  [arXiv:hep-th/0503036].
  %%CITATION = HEP-TH 0503036;%%







%\cite{Kim:2000ax}
\bibitem{Kim:2000ax}
  J.~Y.~Kim,
  ``Brane inflation in tachyonic and non-tachyonic type 0B string theories,''
  Phys.\ Rev.\ D {\bf 63}, 045014 (2001)
  [arXiv:hep-th/0009175].
  %%CITATION = HEP-TH 0009175;%%


%\cite{Papantonopoulos:2000wv}
\bibitem{Papantonopoulos:2000wv}
  E.~Papantonopoulos and I.~Pappa,
  ``Cosmological evolution of a brane universe in a type 0 string
  %background,''
  Phys.\ Rev.\ D {\bf 63}, 103506 (2001)
  [arXiv:hep-th/0010014].
  %%CITATION = HEP-TH 0010014;%%


%\cite{Diamandis:2002zn}
\bibitem{Diamandis:2002zn}
  G.~A.~Diamandis, B.~C.~Georgalas, N.~E.~Mavromatos and E.~Papantonopoulos,
  ``Acceleration of the universe in type-0 non-critical strings,''
  Int.\ J.\ Mod.\ Phys.\ A {\bf 17}, 4567 (2002)
  [arXiv:hep-th/0203241].
  %%CITATION = HEP-TH 0203241;%%



%\cite{Jeong:2005py}
%\bibitem{Jeong:2005py}
 % D.~H.~Jeong and J.~Y.~Kim,
 % ``Mirage cosmology with an unstable probe D3-brane,''
 % arXiv:hep-th/0502130.
  %%CITATION = HEP-TH 0502130;%%

%\cite{Jeong:2005py}
\bibitem{Jeong:2005py}
  D.~H.~Jeong and J.~Y.~Kim,
  ``Mirage cosmology with an unstable probe D3-brane,''
  Phys.\ Rev.\ D {\bf 72}, 087503 (2005)
  [arXiv:hep-th/0502130].
  %%CITATION = HEP-TH 0502130;%%





%\cite{Berkooz:2005ym}
\bibitem{Berkooz:2005ym}
  M.~Berkooz, Z.~Komargodski, D.~Reichmann and V.~Shpitalnik,
  %``Flow of geometries and instantons on the null orbifold,''
  arXiv:hep-th/0507067.
  %%CITATION = HEP-TH 0507067;%%





%\cite{Sen:2003tm}
\bibitem{Sen:2003tm}
  A.~Sen,
  ``Dirac-Born-Infeld action on the tachyon kink and vortex,''
  Phys.\ Rev.\ D {\bf 68}, 066008 (2003)
  [arXiv:hep-th/0303057].
  %%CITATION = HEP-TH 0303057;%%


%\cite{Barnaby:2004dp}
\bibitem{Barnaby:2004dp}
  N.~Barnaby and J.~M.~Cline,
  ``Creating the universe from brane-antibrane annihilation,''
  Phys.\ Rev.\ D {\bf 70}, 023506 (2004)
  [arXiv:hep-th/0403223].
  %%CITATION = HEP-TH 0403223;%%

%\cite{Cline:2005zy}
\bibitem{Cline:2005zy}
  J.~M.~Cline,
  ``Inflation from string theory,''
  arXiv:hep-th/0501179.
  %%CITATION = HEP-TH 0501179;%%

%\cite{Hellerman:2001yi}
\bibitem{Hellerman:2001yi}
  S.~Hellerman, N.~Kaloper and L.~Susskind,
  ``String theory and quintessence,''
  JHEP {\bf 0106}, 003 (2001)
  [arXiv:hep-th/0104180].
  %%CITATION = HEP-TH 0104180;%%

%\cite{Fischler:2001yj}
\bibitem{Fischler:2001yj}
  W.~Fischler, A.~Kashani-Poor, R.~McNees and S.~Paban,
  ``The acceleration of the universe, a challenge for string theory,''
  JHEP {\bf 0107}, 003 (2001)
  [arXiv:hep-th/0104181].
  %%CITATION = HEP-TH 0104181;%%



\end{thebibliography}
\end{document}